\documentclass[aps,pra,preprint]{revtex4}
\usepackage{latexsym}
\usepackage{bbm}
\usepackage{graphics}
\usepackage{here}
\usepackage{amsmath}
\usepackage{amssymb}
\usepackage{graphicx}
\usepackage{graphics}

\begin{document}

\newcommand{\beq}{\begin{equation}}
\newcommand{\eeq}{\end{equation}}
\newcommand{\bea}{\begin{eqnarray}}
\newcommand{\eea}{\end{eqnarray}}
\newcommand{\bra}[1]{\left\langle #1 \right\vert}
\newcommand{\ket}[1]{\left\vert #1 \right\rangle}
\newcommand{\al}{\alpha}
\newcommand{\ua}{\uparrow}
\newcommand{\da}{\downarrow}
\newcommand{\E}{\hat{\cal E}}
\newcommand{\R}{\hat{R}}
\newcommand{\F}{\cal F}
\newcommand{\kr}{\hat{E}}
\newcommand{\den}{\hat{\rho}}
\newcommand{\p}{\hat{P}}
\newcommand{\sgzero}{H}
\newcommand{\sgone}{V}
\newcommand{\zero}{\ket{\sgzero}}
\newcommand{\one}{\ket{\sgone}}
\newcommand{\ground}{\ket{0}}
\newcommand{\brazero}{\bra{\sgzero}}
\newcommand{\braone}{\bra{\sgone}}
\newcommand{\braground}{\bra{0}}
\newcommand{\zerol}{\ket{0_L}}
\newcommand{\onel}{\ket{1_L}}
\newcommand{\brazerol}{\bra{0_L}}
\newcommand{\braonel}{\bra{1_L}}
\newcommand{\gen}{\ket{Q}}
\newcommand{\diss}{\gamma}
\newcommand{\alp}{\alpha}
\newcommand{\ang}{\phi}
\newcommand{\syndrome}{\hat{S}}
\newcommand{\fid}{{\cal F}}
\newcommand{\x}{\hat{\sigma}_x}
\newcommand{\y}{\hat{\sigma}_y}
\newcommand{\z}{\hat{\sigma}_x}
\newcommand{\n}{n}
\newcommand{\key}{k}
\newcommand{\tee}{t}
\newcommand{\prob}{P}
\newcommand{\eff}{{\cal E}}
\newcommand{\lbracket}{[[}
\newcommand{\rbracket}{]]}

\title{On the efficiency of nondegenerate quantum error correction codes for Pauli channels}

\author{Gunnar Bj\"{o}rk}
\affiliation{School of Information and Communication Technology,
Royal Institute of Technology (KTH), Electrum 229, SE-164 40 Kista,
Sweden} \email{gbjork@kth.se} \homepage{http://www.quantum.se}

\author{Jonas Alml\"{o}f}
\affiliation{School of Information and Communication Technology,
Royal Institute of Technology (KTH), Electrum 229, SE-164 40 Kista,
Sweden}

\author{Isabel Sainz\footnote{Now at Departamento de F\'{\i}sica, Universidad de Guadalajara, 44420~Guadalajara, Jalisco, Mexico}}
\affiliation{School of Information and Communication Technology,
Royal Institute of Technology (KTH), Electrum 229, SE-164 40 Kista,
Sweden}

\begin{abstract}
We examine the efficiency of pure, nondegenerate quantum-error
correction-codes for Pauli channels. Specifically, we investigate if
correction of multiple errors in a block is more efficient than
using a code that only corrects one error per block. Block coding
with multiple-error correction cannot increase the efficiency when
the qubit error-probability is below a certain value and the code
size fixed. More surprisingly, existing multiple-error correction codes with
a code length $\leq 256$ qubits have lower efficiency than the optimal single-error correcting codes for any value
of the qubit error-probability. We also investigate how efficient
various proposed nondegenerate single-error correcting codes are
compared to the limit set by the code redundancy and by the necessary conditions for hypothetically
existing nondegenerate codes. We find that existing codes are close
to optimal.
\end{abstract}

\maketitle

\section{Introduction}

Quantum computers hold great promise for efficient computing, at
least for certain classes of problems \cite{nielsen}. However,
similarly to ordinary computers, quantum computers are subject to
noise (unwanted interaction with the environment). Hence, the state
of a quantum computer needs to be monitored and subject to
``restoring forces'' to keep the computation on track. Fortunately,
it has been shown that it is possible to implement such forces,
e.g., quantum error correction, that will ensure that errors do not
lead to computational failures if the qubit error-probability is
kept within certain limits.

Quantum-error correcting codes were discovered by Shor \cite{shor} and by Steane  \cite{steane,steane2}. Soon thereafter a more
conceptual understanding of quantum-error correction-codes developed
\cite{calderbank,steane5,bennett,knill}, and recently a generalized
approach to different kinds of error control, including
decoherence-free subspaces has been developed \cite{kribs}. A large
number of quantum-error correcting codes have been proposed
\cite{ekert,paz,qec,steane4,gottesman,gottesman2,cleve,plenio97,leung,rains2,grassl,calderbank2,steane3,rains,rains3,braunstein,rains4,ashikhmin3,rains5,fletcherN,smolin}.
Recently, work has also been undertaken to develop algorithms for
code optimization \cite{grassl2,reimpell,yamamoto,fletcher1}.
Quantum-error correction-coding is based on a mapping of $\key$
logical qubits onto $\n > \key$ physical qubits. If such a code can
correct up to $\tee$ qubit errors of some restricted class, then the
code is denoted a $\lbracket \n,\key,2 \tee +1\rbracket$ code. The
parameter $2 \tee +1 = d$ is the codeword space distance, and the
distance will have to be $2 \tee +1$ to uniquely identify every
error, as a distance of $2 \tee$ would lead to different errors
resulting in the same state. The numbers $ \n,\key,\tee $ are of
course not independent, but bounds for codes maximizing the ratio
$\key/\n$ for a given $\tee$ have been derived
\cite{calderbank,calderbank2,ashikhmin,ashikhmin2} It has, e.g.,
been established that error correcting codes exist with the
asymptotic rate \cite{calderbank} \beq \key/\n=1-2 H_2(2 \tee/\n),
\label{Eq: Calderbank limit}\eeq where $H_2$ is the binary entropy
function $H_2(p)=-p \log_2(p)-(1-p) \log_2(1-p)$. Hence, the
redundancy (overhead) is rather small for long codes.

In this work, we will consider errors induced by so-called Pauli
channels. The error operators in this case either flip the qubit
value $\ket{0} \leftrightarrow \ket{1}$, flip the qubit phase
$\alpha \ket{0} + \beta \ket{1} \leftrightarrow \alpha \ket{0} -
\beta \ket{1}$, or do a combination of both operations. The errors
can be operationally described by the Pauli operators, hence the
name. For simplicity (and quite realistically) we shall assume that
each qubit in the code are affected by each of these errors
independently, each with a probability $p/3$. Hence, we shall
consider a depolarizing channel, which is a special case of a Pauli
channel. However, the codes we shall discuss can handle any Pauli
channel, although if the possible errors did not occur with the same
probability, somewhat more efficient codes could be constructed
\cite{sarvepalli}. However, we are confident that an analysis of
such codes would qualitatively lead to the same conclusions.

Originally, it was thought that every error must be uniquely
identifiable by the code's error syndrome vector, that is the ensuing vector after at most $\tee$ errors have occurred. If for every error (up to $\tee$ errors) the resulting syndrome vectors are all different, the code is called nondegenerate. If, in addition, all these vectors are
mutually orthogonal, the code is called pure. However, in 1996 codes
were discovered where some errors led to the same syndrome
\cite{shor96,plenio97}. Such a code is called a degenerate code.
Since, a number of different suggestions for degenerate codes have
been put forward, and it has been shown that they provide a higher
communication rate than nondegenerate codes \cite{smith}, at least for Pauli channels. However, the best such codes use concatenation
which is resource demanding. In this work we will therefore take a
step back and analyze pure, nondegenerate codes, and specifically
try to answer the question whether or not it ''pays'' to correct
more than one error per codeword.

\section{The quantum Hamming bound and other restrictions}

A nondegenerate quantum-error correcting code is constructed in such
a way that every detectable error results in a unique syndrome. For
a Pauli channel each qubit can be affected by three different
errors, a bit flip, a phase flip, or both. If $\key$ logical qubits
are coded onto $\n$ physical qubits, and up to $\tee$ errors are to
be uniquely detected, then the quantum Hamming bound must be
fulfilled \cite{ekert}: \beq 2^{\n - \key}
\geq \sum_{i=0}^\tee 3^i  \left(%
\begin{array}{c}
  \n \\
  i \\
\end{array}
\right). \label{Eq:Hamming}\eeq This bound gives a necessary
condition on the size of the syndrome Hilbert space to
accommodate an orthogonal vector for each detectable error. However,
there is no guarantee that a code can be found for every triplet
$\n,\key,\tee$ that satisfies the inequality. In the following we shall
use the designation ``hypothetical code'' for a code labeled
$\lbracket\n,\key,2 \tee +1\rbracket$ where the triplet fulfills the
quantum Hamming bound and other known bounds (see below). Such triplets $\lbracket\n,\key,2 \tee +1
\rbracket$ for which a code is known to exist we shall call existing
codes. Hence, the set of existing codes is a subset of the set of hypothetical codes. For certain triplets, such as $\lbracket 5,1,3 \rbracket$ the
Hamming bound is fulfilled with equality. If a code exists for such
a triplet, then the code is called a perfect code
\cite{bennett,paz}. It is also known that codes can be constructed
for triplets within the Gilbert-Varshamov bound \beq 2^{\n - \key}
\leq \sum_{i=0}^{2 \tee} 3^i  \left(%
\begin{array}{c}
  \n \\
  i \\
\end{array}
\right). \eeq That is, the Gilbert-Varshamov bound gives a lower
limit for the needed ratio $\key/\n$ for a given $\tee$, just like
the bound (\ref{Eq: Calderbank limit}). These bounds, along with
several others \cite{calderbank2}, hence give sufficient bounds.
However, they tend to give ratios $\key/\n$ quite a bit below the
ratios $\key/\n$ achievable with the best existing codes.

The quantum Hamming bound is not the only necessary bound. Knill and
Laflamme \cite{knill} have shown that all quantum codes must fulfill
the quantum Singleton bound \beq \key \leq \n - 4 \tee . \label{Eq
knill} \eeq A similar and related bound is derived in
\cite{calderbank2}. It is shown that a pure code needs to
satisfy \beq \key \leq \n - 2 d +2 , \label{Eq: calderbank} \eeq
where $d$ is the code-word distance. Since an error-correcting,
nondegenerate code requires $d = 2 \tee +1$,  the bounds (\ref{Eq
knill}) and (\ref{Eq: calderbank}) coincide for this case.
Unfortunately these are rather lax bounds. In fact, any code
satisfying the quantum Hamming bound (\ref{Eq:Hamming}) will also
satisfy the bound (\ref{Eq knill}).

A stricter but unfortunately more complicated bound for
Pauli-channel codes was derived in \cite{calderbank2} (as Theorem
21). The bound is expressed in a set of equations, whose solution
can typically only be found through a computer search via linear
programming. The authors of \cite{calderbank2} have searched through
all the possible codes for $\n \leq 30$ and tabulated possible
values for $\key$ and $d$. In Ref. \cite{Grassl3}, an updated table including codes up to $\n \leq 128$ can be found. We have used the tabulated bounds whenever
they are stricter than the Hamming bound for any hypothetical code
with $\n \leq 128$, but above this value we have used the
quantum Hamming bound and in some cases extrapolated values. Hence, the reader should be warned that with
all likelihood, some codes we have hypothetically assumed to exist
may violate the stricter Theorem 21 in \cite{calderbank2}.

\section{Efficiency measures}
\label{Sec: Efficiency}
If one has a block of $m \gg 1$ logical qubits one has many
possibilities to code the block. One could code each qubit
separately, using some $\lbracket \n_1,1,2 \tee_1 +1\rbracket$ code,
one could code them pairwise using some $\lbracket \n_2,2,2 \tee_2
+1 \rbracket$, code, or in the extreme case, code the whole block
using an $\lbracket \n_m,m,2 \tee_m +1\rbracket$ code. In this paper
we shall specifically study how nondegenerate codes' efficiency vary with the correction depth $\tee$ under the restriction that $\n \leq 256$. We shall always consider the asymptotic rate, that is, when the number of qubits $m$ one wants to transmit fulfills $m \gg 256$.

There are several possibilities to define the efficiency of a code.
The best method, from an information-theoretic viewpoint, would be
to define the efficiency as the worst case, or the average, over all
$\n$-qubit density operators (possibly with the restriction to pure
density operators) of \beq \frac{1}{\n}  I_c (\hat{R} \circ \hat {
E} \hat{\rho}_\n, \hat{\rho}_\n), \label{Eq best definition} \eeq
where $I_c$ is the mutual information, $\hat{ E}$ denotes the
statistical error operator, and $\hat{ R}$ represents the syndrome
measurement and recovery operator. However, such a measure is
difficult to compute. It, e.g., requires that one makes
{\it a priori} assumptions about the statistical distribution of the
logical qubit block and then computes the average mutual information
for this weighted ensemble. Such a calculation would be
computationally ``heavy'' even for rather small $\key$ or $\n$.

Another measure would be to look for a worst case scenario of
transmitting an entangled qubit-block. One could then take the
ratio between the original entanglement and the residual
entanglement after coding, Pauli errors, syndrome measurement, and
error correction and multiply with $\key/\n$. Here one would be up
to the daunting task of first defining a sensible quantitative
measure of multiparty entanglement, and then search over the
$2^m$-dimensional Hilbert space for the worst case. Methods for
doing this even for a rather small number of logical qubits (say, $k>5$) are
presently missing.

Yet another measure is the computed worst case, or average, fidelity
between the original logical qubits and the qubits after coding
them, introducing qubit errors with probability $p$ per qubit,
making a syndrome measurement, and error correcting the ensuing
states. One should subsequently multiply this average fidelity with
the ratio $\key/\n$. A good code would give a high fidelity for an
$\n$ not greatly exceeding $\key$. Again, it will be difficult to
compute both the worst case and the average fidelity.

We have opted to to base our efficiency measure on the lower bound of the probability $\prob$
for correctly coding, transmitting, and recovering each block of
$\key$ logical qubits. Suppose that with probability $\prob$ no more that $\tee$ errors occurred in an $\n$ qubit string, coded by a $\lbracket\n,\key,2 \tee +1\rbracket$ code, where, under the assumption of independent errors $\prob$ is given by \beq
\prob = \sum_{i=0}^\tee (1-p)^{\n-i} p^i \left(%
\begin{array}{c}
  \n \\
  i \\
\end{array}
\right). \label{Eq: pcorr}\eeq In this case the code will enable us to restore the string to its correct state. If more than $t$ errors occurred, then in spite of the code, we cannot correct the string, so on average it is fair to assume that we cannot do better than guessing the correct state. In the (logical) Hilbert space of dimension $k$, this would lead to an average fidelity of $2^{-\key}$. Hence, the average fidelity after correction would thus be $\approx \prob \cdot 1 + (1-\prob)/2^\key$. However, for long codes and small qubit error probabilities $p$, both $1-\prob$ and $2^{-\key}$ are small, or even very small, compared to $\prob$, so $\prob$ would give a lower bound to the fidelity, and moreover, be a rather close estimate. (We shall return to this topic in Sec. \ref{sec: Fidelity}.) Our wish is to transmit the maximum information per used physical qubit. Hence, our efficiency measure will be defined \beq \eff = \frac{\prob \key}{\n}. \label{Eq: efficiency}\eeq which essentially tells us how many logical qubits per physical qubit the code can transmit at a certain error probability $p$ per qubit. Inserting Eq. (\ref{Eq: pcorr}) in
Eq. (\ref{Eq: efficiency}) gives our efficiency measure explicitly. In accordance with the law of large numbers, if the number of logical qubits to be transmitted is $K \gg k$, one can expect to receive $\prob K$ correct qubits if one is willing to transmit $\prob K/\eff$ physical qubits.

\section{Multiple error correction}

\begin{center}
\begin{figure}
\includegraphics[width=0.90\textwidth]{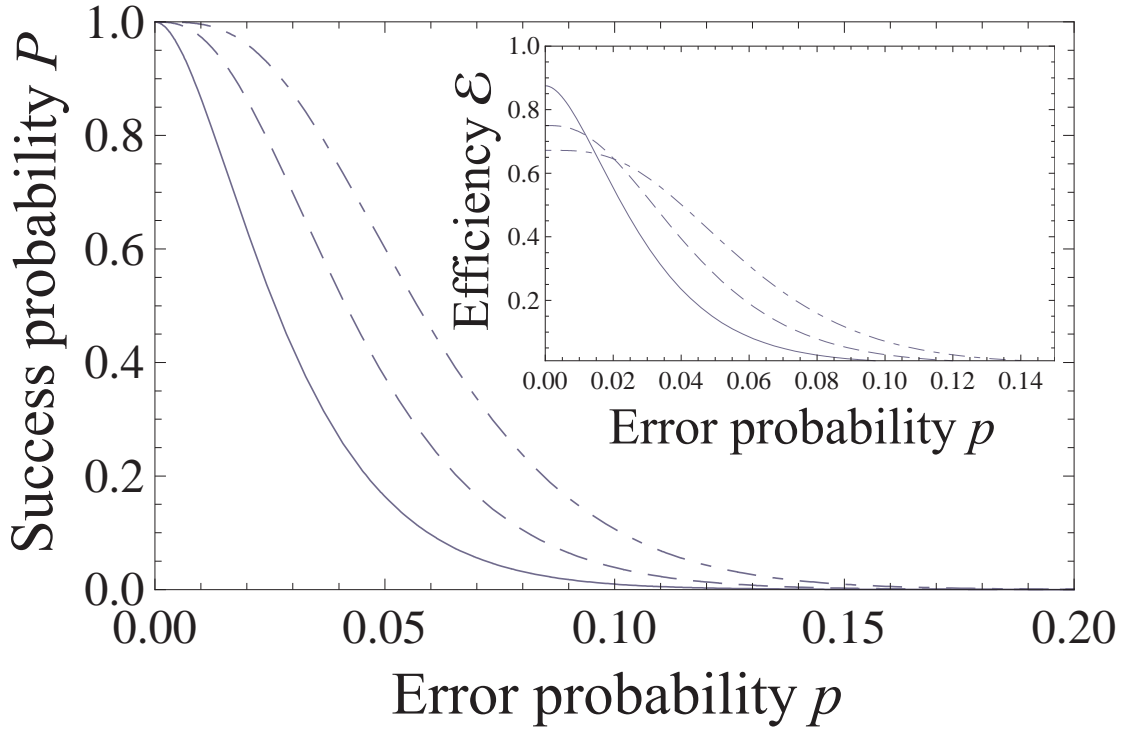}

\caption{The probability that the error-corrected state is identical
to the original state for different codes. The codes are assumed to
have the parameters $\lbracket 64,56,3 \rbracket$ (solid),
$\lbracket 64,48,5 \rbracket$ (dashed), and $\lbracket 64,43,7
\rbracket$ (dot-dashed). Inset the codes' efficiency
$\eff$ is plotted.}\label{fig1}
\end{figure}
\end{center}

In this section the advantages and disadvantages with multiple error
correcting codes will be discussed. We shall mostly take
existing codes as our examples for simplicity. Below we
shall see that existing codes come very close to hypothetical codes
in performance, so possible gains with codes invented in the future will be small. By the way of
example we shall initially study codes 64 qubits long. Using the
Hamming bound (\ref{Eq:Hamming}) and \cite{Grassl3}, one can show that the codes
$\lbracket 64,56,3 \rbracket$, $\lbracket 64,48,5 \rbracket$ exist, and that
$\lbracket 64,43,7 \rbracket$ is a hypothetical code. Actually, the Hamming bound (\ref{Eq:Hamming})
also allows, e.g., the code $\lbracket 64,49,5 \rbracket$, but the more restrictive conditions applied in
\cite{Grassl3} show that such code, in fact, does not exist. In
Fig. \ref{fig1} we have plotted the probability $\prob$ of
transmitting the state correctly, v.s. the single qubit error
probability $p$. A code that corrects single qubit errors has a $1 -
{\cal O}(p^2)$ behavior whereas a code correcting $\tee$ errors will
scale as $1 - {\cal O}(p^{\tee + 1})$ close to $p=0$. Hence as
$\tee$ grows, the code becomes more and more tolerant to errors
while for a fixed code length it can code fewer and fewer logical
qubits, decreasing the efficiency when $p$ is close to zero. This makes sense as strings with few multiple errors will not gain much from multiple error correction.

In Fig. (\ref{fig1}) we have also plotted, as an inset, the
efficiency $\eff$ v.s. the the single qubit error probability $p$.
Here we see that as expected, for a fixed code length the smaller
the error correction depth $\tee$, the more efficient the code is
close to $p = 0$. The codes with larger error correction depth
$\tee$ are only more efficient than the codes with smaller $\tee$
when the error probability $p$ is substantial.

\begin{center}
\begin{figure}
\includegraphics[width=0.90\textwidth]{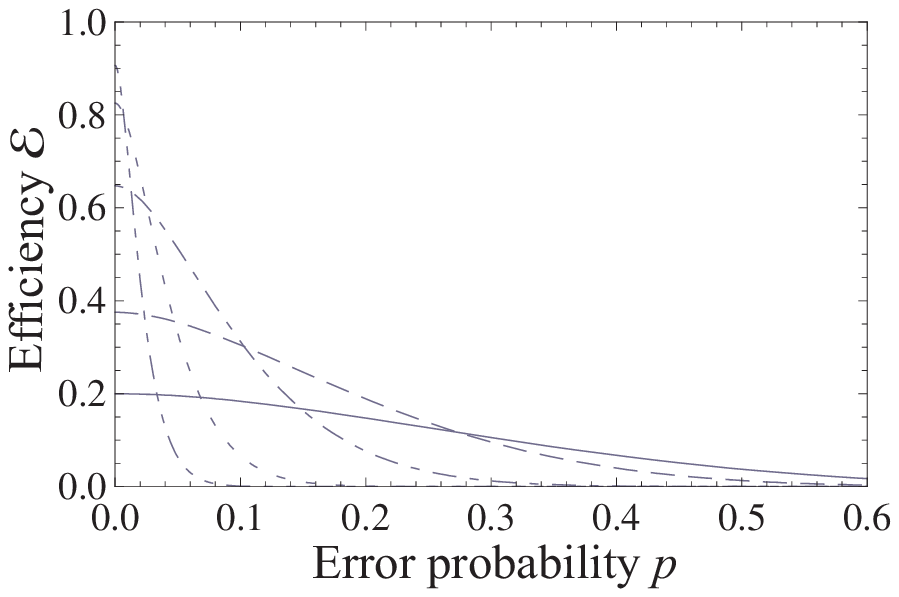}

\caption{The efficiency for codes with assumed parameters $\lbracket
5,1,3 \rbracket$ (solid), $\lbracket 8,3,3 \rbracket$ (dashed),
$\lbracket 17,11,3 \rbracket$ (dot-dashed), $\lbracket 40,33,3
\rbracket$ (small dashed), and $\lbracket 85,77,3 \rbracket$
(dot-dot-dashed).}\label{fig2}
\end{figure}
\end{center}

It has been known for some time now that efficient nondegenerate
quantum codes exist for small errors. The efficiency increases with
increased block length $\key$ for a fixed error correcting depth
$\tee$. In Fig. \ref{fig2} we have plotted the efficiency $\eff$
v.s. the single qubit error probability $p$ for existing
codes correcting a single error, where, moreover, the codes $\lbracket
5,1,3 \rbracket$ and $\lbracket 85,77,3 \rbracket$ are perfect codes. This shows that long codes have the best efficiency, but only for a small range of error probabilities
$p$ close to zero. The envelope of such a set of functions is actually
the real point of interest, because it gives a bound for the
efficiency of any depolarizing-channel, pure, nondegenerate code that is
correcting only a single error per block.

\begin{center}
\begin{figure}
\includegraphics[width=0.90\textwidth]{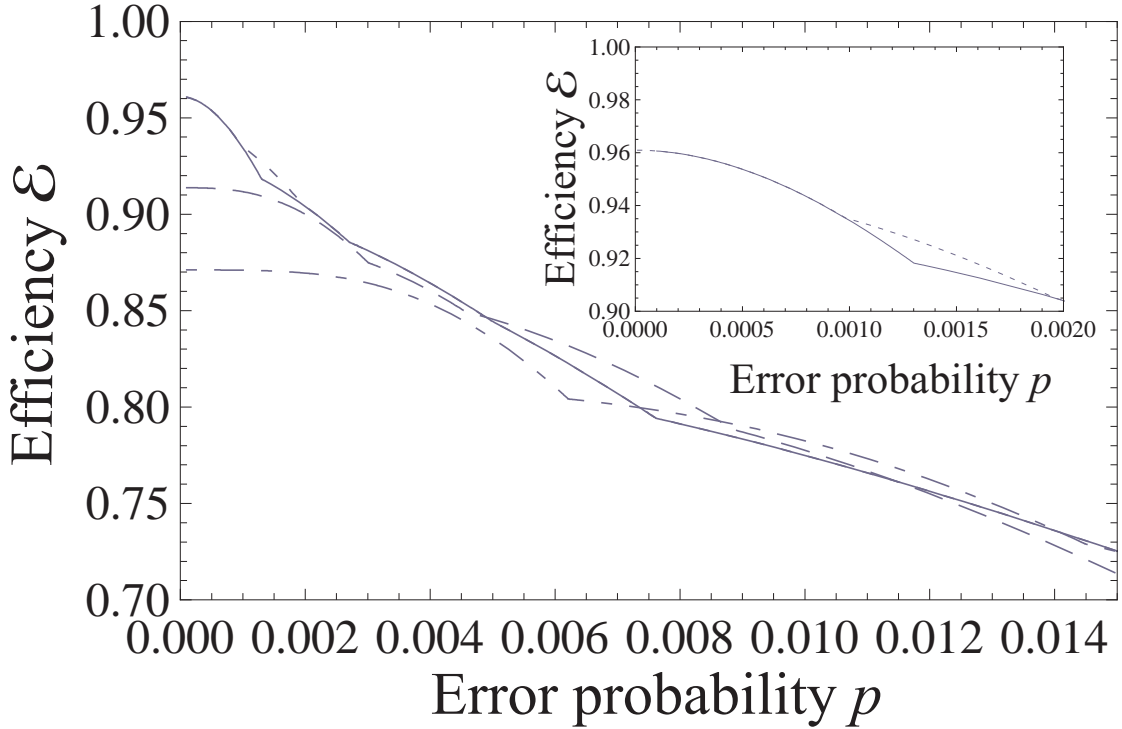}

\caption{The maximum efficiency for hypothetical codes with assumed
parameters $\tee=1$ (short-dashed), $\tee=2$ (dashed), $\tee=3$
(dot-dashed), and the most efficient existing codes with $\tee=1$
(solid). An enlargement of the plotted efficiency of existing and
hypothetical codes with $\tee=1$ in the only region they differ is
inset for clarity.}\label{fig3}
\end{figure}
\end{center}

Let us now compare the efficiency envelope functions of existing and
hypothetical codes with an error correction depth of 1, 2, and 3,
all having $\n \leq 256$ in Fig. \ref{fig3}. To
obtain these envelopes we plot the largest efficiency of the
hypothetical codes fulfilling (\ref{Eq:Hamming}) and for $\n \leq
128$ also the tabulated bounds in \cite{Grassl3}. It is trivially
true that very close to $p=0$ the codes with the largest ratio
$\key/\n$ for a given difference $\n - \key$ will have the highest
efficiency. For larger $p$ this no longer holds true, so one needs
to search trough a large number of hypothetical codes to make sure
that one has found the most efficient code for any given $p$. A
guiding principle in this search is that codes coming close to the
Hamming bound are efficient as they use the code's redundancy nearly
optimally. The most efficient (each for some range of probabilities
$p$), existing, single error correcting codes with $\tee=1$ and $\n
\leq 256$ are $\lbracket 5,1,3 \rbracket$ (perfect), $\lbracket
8,3,3 \rbracket$, $\lbracket 15,9,3 \rbracket$, $\lbracket 16,10,3
\rbracket$, $\lbracket 17,11,3 \rbracket$, $\lbracket 21,15,3
\rbracket$ (perfect), $\lbracket 40,33,3 \rbracket$, $\lbracket
74,66,3 \rbracket$, $\lbracket 85,77,3 \rbracket$ (perfect),
$\lbracket 128,119,3 \rbracket$, and $\lbracket 256,246,3
\rbracket$. The implementation of these codes (and the codes below)
can be found in \cite{calderbank2,Grassl3}, where the latter
reference is the more extensive. The only hypothetical $\tee=1$ code
we have found that could possibly beat any of these codes (but only
in a small range of $p$) is a $\lbracket 170,161,3 \rbracket$ code, as shown in the Fig. \ref{fig3} inset.

The $\n \leq 256$ codes with $\tee=2$ which have the highest
efficiencies are $\lbracket 11,1,5 \rbracket$, $\lbracket 16,4,5
\rbracket$, $\lbracket 18,6,5 \rbracket$, $\lbracket 27,13,5
\rbracket$, $\lbracket 30,16,5 \rbracket$, $\lbracket 35,20,5
\rbracket$, $\lbracket 58,42,5 \rbracket$, $\lbracket 70,54,5
\rbracket$, $\lbracket 128,110,5 \rbracket$ and $\lbracket 256,231,5
\rbracket$. Hypothetically, codes with the following parameters will
be even more efficient: $\lbracket 14,3,5 \rbracket$, $\lbracket
16,5,5 \rbracket$, $\lbracket 17,6,5 \rbracket$, $\lbracket 27,15,5
\rbracket$, $\lbracket 39,26,5 \rbracket$, $\lbracket 83,68,5
\rbracket$, $\lbracket 118,102,5 \rbracket$, $\lbracket 170,151,5
\rbracket$, and $\lbracket 256,233,5 \rbracket$. The last two codes
need an additional comment. The Hamming bound allows $\lbracket
170,153,5 \rbracket$ and $\lbracket 256,236,5 \rbracket$ codes, but
an analysis of both existing and hypothetical codes with $\n \leq
128$ shows that neither set of codes come close to saturating the
Hamming bound. Both sets of codes have a nearly linear relationship
between $\key$ and $\n$, so we have been a little bit conservative
and estimated the last two hypothetical codes by linear
extrapolation of the rest of the ``hypothetical'' set, and used
these extrapolated parameters in plotting Fig. \ref{fig3}.

The most efficient, existing $\tee=3$ codes are $\lbracket 17,1,7
\rbracket$, $\lbracket 25,5,7 \rbracket$, $\lbracket 35,13,7
\rbracket$, $\lbracket 42,20,7 \rbracket$, $\lbracket 64,38,7
\rbracket$, $\lbracket 113,85,7 \rbracket$, $\lbracket 128,98,7
\rbracket$, and $\lbracket 255,215,7 \rbracket$. The hypothetical
codes with higher efficiency are $\lbracket 20,3,7 \rbracket$,
$\lbracket 22,5,7 \rbracket$, $\lbracket 28,11,7 \rbracket$,
$\lbracket 36,18,7 \rbracket$, $\lbracket 59,39,7 \rbracket$,
$\lbracket 94,72,7 \rbracket$, $\lbracket 121,98,7 \rbracket$, and
$\lbracket 256,223,7 \rbracket$, where the last code again is
inferred through linear extrapolation from the $\n \leq 128$ codes
in the corresponding set, whereas the quantum Hamming bound permits
the code $\lbracket 256,229,7 \rbracket$.

In Fig. \ref{fig3} one sees that, as expected,
for sufficiently small errors $p$, the single-error correcting code
is most efficient, whereas for larger $p$ it
is conceivable that $t=2$ or $t=3$ codes could be more efficient.
(Remember that so far we are considering hypothetical codes for
$\tee>1$.)

It is interesting to see for what interval of qubit error
probabilities $0 < p \leq p_c$ it is impossible to find $\tee>1$
codes that outperform the $\tee = 1 $ codes, given $\ \leq 256$. To
find the interval we need to compute the qubit error probability
$p_c$ for which the efficiency of the most efficient, existing
$\lbracket \n, \key_1,3 \rbracket$ and the best hypothetical
$\lbracket \n, \key_2,5 \rbracket$ codes are equal. To give an
analytical estimate of the range $0 \leq p \leq p_c$ where single
error correcting codes are the most efficient we use Eqs. (\ref{Eq:
pcorr}) and (\ref{Eq: efficiency}) and set $\eff( \n, \key_1,3,p_c )
= \eff( \n, \key_2,5,p_c)$. If we expand the ensuing expression to
second order in $p_c$ and solve the equation, we find that \beq p_c
\approx \sqrt{\frac{2(\key_1 - \key_2)}{\key_1 \n (\n -1)}}
.\label{Eq: p crossover}\eeq Since the difference $\key_1 - \key_2$
grows only very slowly with $\n$, but $\key_1$ grows approximately
linearly with $\n$, we conclude that for $0 < p \leq p_c$, where
$p_c \sim n^{-3/2}$, there will exist a nondegenerate, length $\n$,
single qubit correcting code with higher efficiency than any length
$\n$, two-qubit correcting code. Using the values of $\n$, $\key_1$
for the existing code $\lbracket 256,246,3 \rbracket$ and $\key_2$
for the hypothetical code $\lbracket 256,233,5 \rbracket$, we find
that the range of error probabilities where the former code
outperforms the latter is $0 < p \leq 0.0013$. A more conservative
estimate through the quantum Hamming bound, which allows a
$\lbracket 256,237,5 \rbracket$ code, gives the bound $0 < p \leq
0.0011$. This range of probabilities is surprisingly large.

However, as seen in Fig. \ref{fig3}, before the respective curves for the $\lbracket 256,246,3 \rbracket$ and $\lbracket 256,233,5 \rbracket$ codes above cross, the  existing $\lbracket 128,119,3\rbracket$ code becomes more efficient. Therefore, it is only for approximately $p \geq 0.0025$ any (so far hypothetical) $\tee >1$ code become more efficient than any (existing) $\tee =1$ code.

\section{How good are existing codes?}

In Fig. \ref{fig3}, inset, the efficiency of the most efficient, existing,
nondegenerate codes correcting one-qubit errors and the efficiency
of the most efficient similar hypothetical codes is plotted. As mentioned above we have found only one hypothetical code that in a small range of error probabilities could outperform the existing codes, so there is little hope for improvement for the $\tee = 1$ codes. One may ask why this is, and the answer is that the existing $\tee = 1$ codes listed in the previous section all lie very close to the quantum Hamming bound which provides the least restrictive necessary condition for nondegenerate codes. E.g., the $\lbracket 256,246,3 \rbracket$ code uses 769 out of the possible 1024 syndrome vectors to correct all single qubit errors. That is, the code uses most of its redundancy to perform the task it is intended for -- to correct the most frequent errors. We shall see below that in the range of error probabilities where a specific code is the most efficient, it is impossible to design more than a marginally more efficient code, even under the most optimistic assumptions. The nondegenerate perfect codes are simply perfect. They use all redundancy for the intended purpose.

\begin{center}
\begin{figure}
\includegraphics[width=0.90\textwidth]{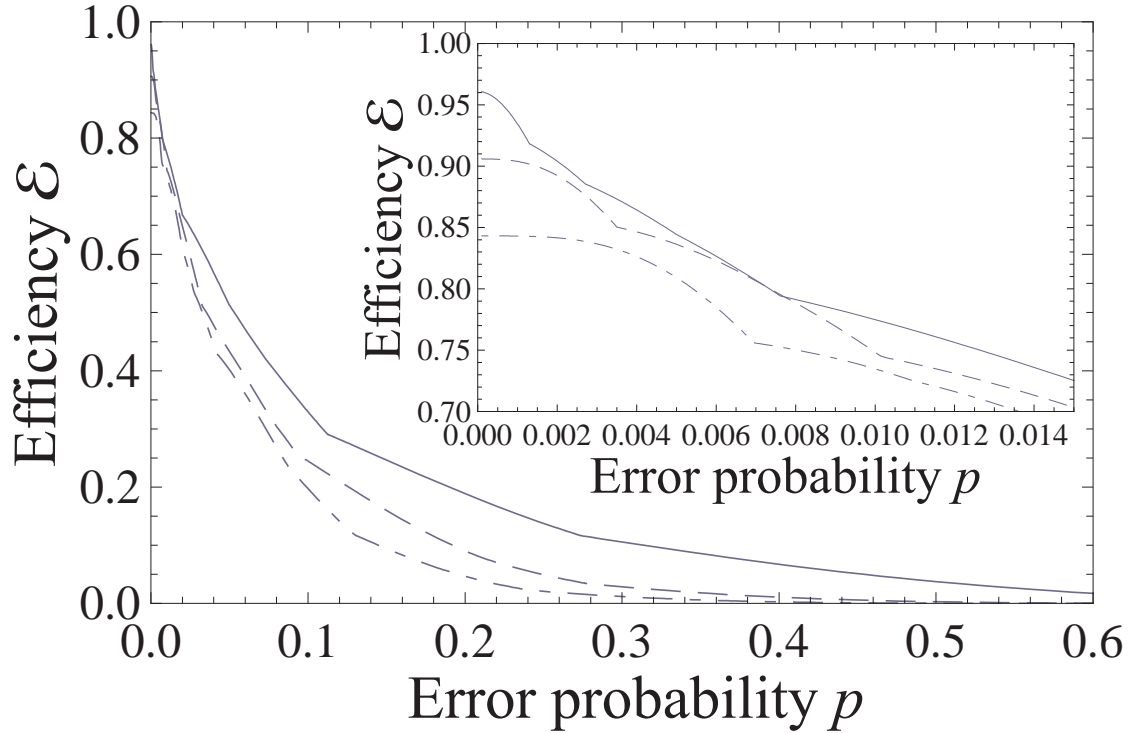}

\caption{The efficiency for some existing codes with $\n \leq 256$
vs. the qubit error probability $p$ for $\tee=1$ (solid), $\tee=2$
(dashed), and $\tee=3$ (dash-dotted).}\label{fig4}
\end{figure}
\end{center}

In Fig. \ref{fig4} we plot the maximum efficiency of $\n \leq 256$
existing nondegenerate codes, listed in the previous section. Whereas it is quite trivial that for
hypothetical codes, there must always be some small range of $p$
were a single-qubit correcting code must be more efficient than any
$\tee>1$ code, given that they all have a maximum code length $\n_{{\rm Max}}$, the result for existing codes is nontrivial. For the
existing $\n \leq 256$ codes we have analyzed, {\em one can always find
among them a $\tee=1$ code that has equal or higher efficiency for any value
of the qubit error probability $p$ than any existing $\tee > 1$
code.} The reason for this counterintuitive result, that contradicts our experience from classical codes, is that the number of errors increases a factor $3^t$ faster for the Pauli channel than for e.g., a classical flip channel. Therefore, the redundancy must also grow much faster for a Pauli-channel code than for a classical bit-flip code. This makes the efficiency smaller for the $t>1$ than for the $t=1$ codes.

\section{Can existing codes be improved?}
 We have seen above that, excluding the three perfect codes, even the optimal codes do not use all possible syndromes. One may then ask how much could be gained if, hypothetically, the whole syndrome vector space could be used for correcting errors. One should remember that quite obviously the most frequent errors should be corrected with the highest priority, so before attempting to correct any double errors, one should correct all single errors etc., and it is on this premise codes are designed. However, in order to argue that existing codes cannot be much better, let us assume for the moment that codes that are pure, nondegenerate and that uses every syndrome to correct errors can be constructed. The number $r$ of ``left over'' syndrome vectors of a $\lbracket \n,\key,2 \tee + 1 \rbracket$ code is
 \beq r = 2^{(\n-\key)} - \sum_{i=0}^\tee 3^i  \left(
\begin{array}{c}
  \n \\
  i \\
\end{array}
\right)\eeq
and only for perfect codes this number is zero. The number $q$ of unique errors of order $\tee + 1$ is \beq 3^{\tee +1}  \left(
\begin{array}{c}
  \n \\
  \tee +1 \\
\end{array}
\right).\eeq
 Hence we can correct a ratio $r/q$ of the $\tee + 1$ order errors. If we corrected all such errors, the success probability $\prob$ would be boosted by the term \beq \left(
\begin{array}{c}
  \n \\
  \tee +1 \\
\end{array}
\right)(1-p)^{(\n-\tee - 1)}p^{(\tee+1)}.\eeq Because the number of leftover syndromes $r$ is smaller than the number of $\tee + 1$ order errors (if not, the code is ill designed as it has sufficient length and redundancy to correct all errors of order $\tee + 1$) we can not, even in the best case, correct all of them, and therefore the additional contribution to the success probability $\prob$ will be \beq \frac{r}{q}\left(
\begin{array}{c}
  \n \\
  \tee +1 \\
\end{array}
\right)(1-p)^{(\n-\tee - 1)}p^{(\tee+1)} = \frac{r}{3^{(\tee + 1)}}(1-p)^{(\n-\tee - 1)}p^{(\tee+1)}\eeq and the additional contribution to the efficiency $\eff$ will be $\key / \n$ times this number.

 \begin{center}
\begin{figure}
\includegraphics[width=0.90\textwidth]{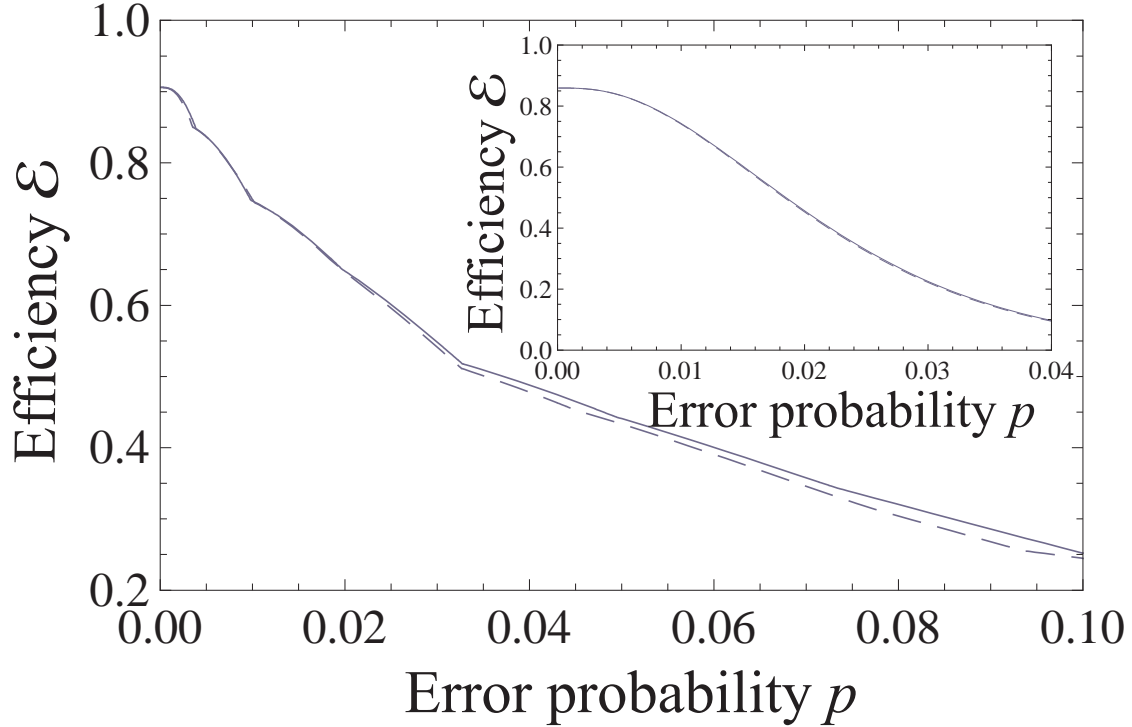}

\caption{The efficiency for existing $t=2$ codes with $\n \leq 256$
(dashed), and for assumed codes of the same lengths that uses every
syndrome for error correction (solid), vs. the qubit error
probability $p$. Inset the performance of the existing $\lbracket
128,110,5 \rbracket$ code (dashed) and an assumed pure,
nondegenerate code with the same $\n$ and $\key$ that uses every
syndrome for error correction (solid) are plotted.}\label{fig5}
\end{figure}
\end{center}

To put these equations in context, let us take the $\lbracket
128,110,5 \rbracket$ code over GF(2*2) \cite{Grassl3} as an example.
The code allows $2^{(128-110)} = 262 144$ syndromes, whereof 1 is
needed to identify the no error case, $3 \cdot 128 = 384$ are needed
to identify all single errors, and $3^2 \cdot 128 \cdot 127 / 2 = 73
152$ are needed to identify all double errors. Remain $r = 262 144
- 73 537 = 188 607$ syndrome vectors that can be used to correct
some of the the $m = 9 217 152$ triple errors. Hence, the efficiency
can be boosted by a term $188 607 \cdot 110 (1-p)^{125} p^3 / (128
\cdot 3^3 ) \approx 6003 (1-p)^{125} p^3$ which looks large, but
which vanishes quickly when $p < (6003)^{-1/3} \approx 0.055$. This
result is intuitive, because if already the probability for single
errors is small, it will certainly not pay to correct triple errors
(which will have a vanishingly small probability to occur). In Fig.
\ref{fig5}, inset, we have plotted the efficiency of the code
$\lbracket 128,110,5 \rbracket$ used as intended to correct up to
double errors (dashed), and the efficiency of a similar length code
where we have assumed that, in addition to all errors up to second
order, we could use the 188 607 ``left over'' syndrome vectors to
identify and correct this number of triple errors. However, as both
the fraction of correctable triple errors is small, and for small
values of $p$ the probability of triple errors is small, the two
curves are almost identical and the difference can hardly been seen
within the resolution of the figure. Moreover, as can be deduced from
Fig. \ref{fig4}, inset, the $\lbracket 128,110,5 \rbracket$ code is no
longer the most efficient code for error probabilities $p > 0.01$,
but the shorter code $\lbracket 70,54,5 \rbracket$ is more
efficient. In Fig. \ref{fig5} we have plotted most efficient
$\tee=2$ existing codes with (dashed) and without (solid) the (with
all likelihood overoptimistic) assumption that all $r$ ``left over''
syndrome vectors can all somehow be used to correct triple errors.
There is little difference
between the two curves, in particular for $p < 0.01$. This suggests that current nondegenerate
codes cannot be improved more than very marginally.

\section{A comment on fidelity, efficiency and mutual information}
\label{sec: Fidelity}

As mentioned in Sec. \ref{Sec: Efficiency} the best  measure of a
code's efficiency would be based on the mutual information between the
original and the corrected qubits. However, the mutual information
is cumbersome to compute, so instead, the fidelity is often used. We
have instead used a measure based on the success probability, and
the reason therefore warrants a short comment.

After coding, and depolarization, a coded qubit block will belong to
one of three categories: (1) It may have suffered $\leq \tee$ errors
and therefore have a syndrome which will be recognized as
correctable. After the recovery operation one will have the desired,
original qubit block. (2) It may have suffered $> \tee$ errors and
have a syndrome that is orthogonal to all syndromes of recoverable
errors. (3) It may have suffered $> \tee$ errors but have a syndrome
that is nonorthogonal to some syndromes of recoverable errors. The syndrome vectors will
then occasionally be recognized as correctable. However, the recovery
operator, intended for syndromes from group (1), will then in general not ``generate'' the desired qubit block but an erroneous one.

If we want to optimize the fidelity we should process the noncorrectable errors belonging both to group (2) and (3) in some manner and produce a proper length qubit block. While it is probably not the optimum operation, we could, e.g., project all these state onto the $\key$ qubit (properly normalized) identity operator. This operator has a fidelity $2^{-\key}$ to any pure initial $\key$ qubit block, so the states in both group (2) and (3) will increase the fidelity if processed this way (if not by much), both from a mathematical and an experimental viewpoint.

On the contrary, neither of the groups (2) or (3) contribute to the success probability, as defined. In an experiment, one should therefore ``throw away'' any state with a syndrome orthogonal to those that signals a correctable error. This will rid us of group (2). Those times the syndromes of states belonging to group (3) are measured as recoverable, one must go ahead with the procedure, for there is no way one can decide if such an outcome is triggered by a state from group from (1) or from group (3). If it is from the latter group, the thus ``recovered'' state will with high likelihood be incorrect but such events will still contribute (very, very marginally) to the measured (but not to the computed) success probability.

However, whereas all states in group (1) contribute positively to the mutual information, the states in group (2) and those events that yields a `nonrecoverable'' syndrome measurement in group (3) will with all likelihood contribute negatively because in general they will result in an incorrect qubit block. Therefore, it seems like the best strategy will be to discard these states. If so, they don't contribute to the mutual information. However, the group (3) states that leads to a measurement event signalling a recoverable error are inevitably going to contribute negatively to the mutual information because they will result in a small portion of erroneous qubit blocks randomly mixed with the successfully recovered qubit blocks from group (1). We believe that this negative contribution will be small, in particular for long codes, for two reasons. The first reason is that even for quite large qubit error probabilities $p$, most states will belong to group (1), that is come out as the desired qubit blocks, provided that one uses a code that has the optimal efficiency for the particular error probability. E.g. the success probability of the code $\lbracket 128, 119, 3\rbracket$ is 0.93 at $p = 0.0035$ (at which point its efficiency is superseded by the $\lbracket 85, 77, 3\rbracket$ code). Hence, ``false corrections'' at this error probability can come only from a (small) fraction of the 7 percent of blocks that have two or more errors. The second reason is that for long codes, the code qubit space dimension $2^\n$ is much larger than the correctable syndrome sub-space with dimension $\leq 2^\key$. Therefore, the overlap between a state in group (1) in the latter space and a state in group (3) in the former but not in the latter space will with all likelihood be very small. ``False corrections'' should hence be very rare for long codes.

From the considerations above we see that some strategies that serve to increase the fidelity actually decreases the mutual information. This is not so with success probability. We conjecture that the estimate of mutual information that can be derived from the success probability is rather close to the actual mutual information, although we have been unable quantify this conjecture. At any rate, it seems like fidelity, in spite of its popularity, is the worst of these three measures from this point of view.

\section{Conclusions}

We have looked at pure, nondegenerate quantum-error correction codes for blocks
of qubits subjected to statistically independent noise in a
depolarizing channel. For a fixed maximum code length $\n$ and a
probability of error below a certain level, the codes that correct
only a single error per block are the most efficient in the ``steady state'', i.e., when the number of qubits to be transmitted is $\gg \n$.  We believe that in analogy with classical error
correction, the hardware-implementation penalty in cost and
complexity will be prohibitive for long codes. Therefore it is
reasonable to compare codes of the same maximum length.

We have subsequently derived an approximate expression for the maximum qubit error
rate where single quantum-error correcting codes will have
the highest efficiency for a given length $\n$ and found that if the
error probability is at most $p \approx 10^{-3}$, then for a code length $\n \leq
256$ it is not efficient to correct more than a single error even if
more efficient nondegenerate, multiple-error correction codes will
be invented. The range is surprisingly wide. Moreover, if the
efficiency of nondegenerate, multiple error correcting codes will
not improve by the invention of new codes, then the single-error
correcting codes are most efficient regardless of the qubit error
probability if one restricts the code length. This came as a surprise for us, but is good news for quantum information technology because as mentioned above, the number of errors, and hence the needed correction apparatus, grows exponentially with the correction depth. For example, the number of single errors per block grows as $3 \n$, whereas the number of double errors grows
as $9 \n(\n-1)/2$.

We have also provided evidence that the existing codes are close to optimal in performance. It hence seems unlikely that work on optimization of the codes considered (pure and nondegenerate for Pauli channels) will lead to more than very marginal improvements.

It is interesting to ask if the conclusions we have drawn above
spill over to other channel models and to degenerate codes. To the
best of our knowledge the answers to these questions are not known.
We should suspect that for nondegenerate codes the result should
hold even for other channels, for the general scaling behavior of
such codes  for $\n$, $\key$, and $\tee$ is similar. How the
efficiency of degenerate codes behaves as a function of error
correction ability is still an open question.

We have also not coupled the codes' efficiency with its fault tolerance, and to the best of our knowledge the efficiency v.s. the fault-tolerance threshold is still an open question. Rather fault-tolerant codes are known \cite{Knill2}, but at the price of a significant overhead (a ratio $\key/\n \ll 1$). To make such a study, it is necessary to couple the qubit error probability to the gate error rate and the gate number, for codes of different correction depth $\tee$. So far, the investigated codes for fault-tolerant quantum computing typically has a rather small number of coded qubits $k$, and efficiency has been sacrificed for achieving a high fault tolerance.

\section*{Acknowledgements}
This work was supported by the Swedish Research Council (VR), the
Swedish Foundation for Strategic Research (SSF), the Swedish
Foundation for International Cooperation in Research and Higher
Education (STINT), and the ECOC 2004 foundation. GB would like to
thank Professor S. Inoue for his generous hospitality at Nihon
University, Dr. M. Grassl for a fruitful correspondence and for giving access to Ref. \cite{Grassl3}, and Dr. J.~S\"{o}derholm for valuable comments.


\begin{thebibliography}{27}

\bibitem{nielsen} For an introduction to quantum computing, see e.g., M.~Nielsen and I.~Chuang, \emph{Quantum Computation and Quantum
Information} (Cambridge University Press, Cambridge, 2000).
\bibitem{shor} P.~W.~Shor, Phys. Rev. A {\bf 52}, R2493 (1995). 

\bibitem{steane} A.~M.~Steane, Phys. Rev. Lett. {\bf 77}, 793
(1996). 

\bibitem{steane2} A.~M.~Steane, Phys. Rev. A {\bf 54}, 4741
(1996). 

\bibitem{calderbank} A. R. Calderbank and P. W. Shor, Phys. Rev. A
{\bf 54}, 1098 (1996). 

\bibitem{steane5} A.~M.~Steane, Proc. Roy. Soc. Lond. A {\bf 452},
2551 (1996). 

\bibitem{bennett}C.~H.~Bennett, D.~P.~DiVincenzo, J.~A.~Smolin, and W.~K.~Wootters, Phys. Rev. A
{\bf 54}, 3824 (1996). 

\bibitem{knill} E.~Knill and R.~Laflamme, Phys. Rev. A {\bf 55},
900 (1997). 

\bibitem{kribs} D Kribs, R. Laflamme, and D. Poulin, Phys. Rev. Lett. {\bf 94},
180501 (2005). 

\bibitem{ekert}  A.~Ekert and C.~Machiavello, Phys. Rev. Lett. {\bf
77}, 2585 (1996). 

\bibitem{paz} R.~Laflamme, C.~Miquel, J.~P.~Paz, and W.~H.~Zurek,
Phys. Rev. Lett. {\bf 77}, 198 (1996). 

\bibitem{qec} D.~Gottesman, Phys. Rev. A {\bf 54}, 1862 (1996);
A.~R.~Calderbank, E.~M.~Rains, P.~W.~Shor, and N.~J.~Sloane, Phys.
Rev. Lett. {\bf 78}, 405 (1997). 

\bibitem{steane4} A. M. Steane, e-print arXiv:quant-ph$\backslash$9802061v2. 

\bibitem{gottesman} D. Gottesman, Phys. Rev. A {\bf 54}, 1862
(1996). 

\bibitem{gottesman2} D. Gottesman, e-print arXiv:quant-ph$\backslash$9607027. 

\bibitem{cleve} R. Cleve and D. Gottesman, Phys. Rev. A {\bf 56}, 76 (1997). 

\bibitem{plenio97} M.~B.~Plenio, V.~Vedral, and P.~L.~Knight, Phys.
Rev. A {\bf 55}, 67 (1997).

\bibitem{leung} D.~W.~Leung, M.~A.~Nielsen, I.~L.~Chuang, and
Y.~Yamamoto, Phys. Rev. A {\bf 56}, 2567 (1997).

\bibitem{rains2} E. M. Rains, R. H. Hardin, P. W. Shor, and N. J. A. Sloane,
Phys. Rev. Lett., {\bf 79}, 953 (1997). 

\bibitem{grassl} M.~Grassl, Th.~Beth, and T.~Pellizzari,
Phys. Rev. A {\bf 56}, 33 (1997).

\bibitem{calderbank2} A. R. Calderbank, E. M. Rains, P. W. Shor, and N. J. A. Sloane,
IEEE Trans. Info. Theory, {\bf 44}, 1369 (1998). 

\bibitem{steane3} A.~M.~Steane, IEEE Trans. Info. Theory {\bf 45}, 1701 (1999). 

\bibitem{rains} E. M. Rains, IEEE Trans. Info. Theory {\bf 45}, 1827
(1999). 

\bibitem{rains3} E. M. Rains, IEEE Trans. Info. Theory {\bf 45}, 266 (1999). 

\bibitem{braunstein} S. L. Braunstein, C. A. Fuchs, D. Gottesman, and H.-K. Lo, IEEE
Trans. Info. Theory, {\bf 46}, 1644 (2000). 

\bibitem{rains4} E. M. Rains, Finite Fields Appl.,
{\bf 6}, 146 (2000). 

\bibitem{ashikhmin3} A. Ashikhmin, S. Litsyn, and M. A. Tsfasman, Phys. Rev. A {\bf 63}, 032311 (2001). 

\bibitem{rains5} E. M. Rains, IEEE Trans. Info. Theory  {\bf 49}, 1261 (2003). 

\bibitem{fletcherN} A.~S.~Fletcher, P.~W.~Shor, and M.~Z.~Win,
Phys. Rev. A {\bf 77}, 012320 (2008).

\bibitem{smolin} J. A. Smolin, G. Smith, and S. Wehner,
Phys. Rev. Lett. {\bf 99}, 1 (2007). 

\bibitem{grassl2} M. Grassl, T. Beth, and T. M. R\"{o}tteler, Int. J. Quantum Inf. {\bf 2}, 55 (2004). 

\bibitem{reimpell} M. Reimpell and R. F. Werner, Phys. Rev. Lett.
{\bf 94}, 080501 (2005).

\bibitem{yamamoto} N. Yamamoto, S. Hara, and K. Tsumura, Phys. Rev.
A {\bf 71}, 022322 (2005).

\bibitem{fletcher1} A.~S.~Fletcher, P.~W.~Shor, and M.~Z.~Win,
Phys. Rev. A {\bf 75}, 012338 (2007).

\bibitem{ashikhmin} A. Ashikhmin and S. Litsyn, IEEE Trans. Info. Theory, {\bf 45}, 1206 (1999). 

\bibitem{ashikhmin2} A. Ashikhmin, A. M. Barg, E. Knill, and S. N. Litsyn, IEEE Trans. Info. Theory, {\bf 46}, 789 (2000). 

\bibitem{sarvepalli} P. K. Sarvepalli, M. R\"{o}tteler, and A. Klappenecker, e-print arXiv:0804.4316. 

\bibitem{shor96} P. W. Shor and J. A. Smolin, e-print arXiv:quant-ph$\backslash$9604006v2. 

\bibitem{smith} G. Smith and J. A. Smolin, Phys. Rev. Lett. {\bf 98}, 030501
(2007). 

\bibitem{Grassl3} M. Grassl, tabulated codes online available at \url{http://www.codetables.de} (2007). Accessed on 2008-10-07.

\bibitem{Knill2} E.~Knill, Nature {\bf 434}, 39 (2005).






\end{thebibliography}
\end{document}